\documentstyle[aps,epsfig,prl,floats]{revtex}
\begin{document}
\draft

\title{Discussion of a spin-cluster model for the low temperature phase of 
	$\alpha$'-NaV$_2$O$_5$}

\author{Simon Trebst\cite{Bonn} and Anirvan Sengupta}
\address{Bell Labs, Lucent Technologies, Murray Hill, NJ 07974}

\twocolumn[\hsize\textwidth\columnwidth\hsize\csname
@twocolumnfalse\endcsname
\date{\today}
\maketitle
\widetext

\begin{abstract}
We discuss magnetic excitations of a spin-cluster model which has been 
suggested to describe the low temperature phase of $\alpha$'-NaV$_2$O$_5$. 
This model fulfills all symmetry criteria proposed by recent x-ray 
investigations.
We find that this model is not able to describe the occurence of two well 
separated magnon lines perpendicular to the ladder direction as observed in 
INS experiments.
We suggest further experimental analysis to generally distinguish between 
models with double reflection or inversion symmetry.
\end{abstract}

]

\narrowtext

The modeling of the low temperature phase of NaV$_2$O$_5$ has initiated a 
discussion of effective spin models for the low-lying magnetic excitations. 
NaV$_2$O$_5$ undergoes a phase transition at T$_c$ = 34K \cite{Isobe:96} 
associated with a lattice distortion, charge ordering and the opening of a 
spin gap. 
In NaV$_2$O$_5$ the V-ions are arranged in staggered ladders along the
crystallographic $b$-axis. While in the high temperatur phase there is only
one V$^{4.5+}$ site, recent x-ray diffraction studies 
\cite{Luedecke:99,deBoer:00,Smaalen:99} suggest that in the low temperature 
phase there are three distinct valence states: on every other ladder one finds 
a zig-zag charge ordering of V$^{4+}$ and V$^{5+}$ valence states while on the 
intermediate ladders one finds rungs with two V$^{4.5+}$ sites.
The structural investigation further indicates that the space group of the
low temperature phase is $Fmm2$. In the $a$-$b$ plane one finds a doubling 
of the unit cell along {\bf a} and {\bf b} as well as mirror planes 
$\perp${\bf a} and $\perp${\bf b}. The latter criterion generally excludes 
models with dimerization along in-line\cite{Thalmeier:98} or zig-zag chains
\cite{Seo:98,Khomskii:98}.

Recent inelastic neutron scattering experiments \cite{Regnault:00} show that 
there are two close-by magnon excitations with a gap of 8.75~meV and 10.65~meV.
Both excitations have a large dispersion along the $b$-axis. The magnetic 
exchange coupling along {\bf b} has been estimated to range between 37.9~meV 
\cite{Yosihama:98} and 60~meV \cite{Regnault:00}.
The dispersion along {\bf a} shows only a weak modulation of about 0.5~meV
which is out of phase for the two well seperated branches.
Raman scattering experiments \cite{Fischer:99} observe three excitations 
below the two magnon continuum which have been interpreted as singlet 
excitations. Remarkably, the lowest excitation has a gap that seems to 
coincide with the gap of the lower branch of the two magnon excitations.

\begin{figure}[h]
 \begin{center} 
 \epsfig{file=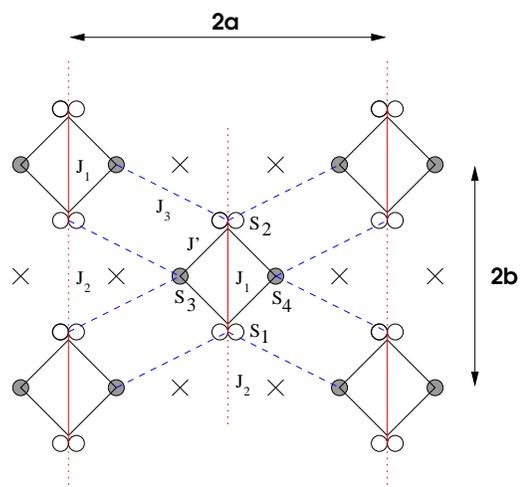, height=6.5cm}
 \caption[]
	{The spin-cluster model. 
	 The filled circles denote V$^{4+}$-ions, the crosses denote 
	 V$^{5+}$-ions and the open circles denote pairs of V$^{4,5+}$-ions.}
 \label{ClusterSpinModel}
 \end{center}
\end{figure}

Based on recent x-ray diffraction experiments in the low temperature phase 
de Boer et al. \cite{deBoer:00} proposed the formation of weakly coupled, 
frustrated spin-clusters along the crystallographic $b$-axis \cite{Marel:00}. 
Each cluster (see Fig. \ref{ClusterSpinModel}) contains six Vanadium atoms 
distributed over three ladders and and an overall number of four unpaired 
electrons which form a singlet ground state. 
This spin-cluster model is the only proposed model that obeys double reflection
symmetry.
\\

A previous theoretical study \cite{Gros:00} addressed the applicability of the
one-dimensional arrangement to model the strong magnon dispersion along the 
crystallographic $b$-axis. 
Using a novel cluster-operator theory as well as exact diagonalization and
DMRG calculations the authors concluded that there is no parameter regime which
would reproduce the observed $b$-axis dispersion.

In this paper, we present a study of the proposed spin-cluster model by means 
of a strong-coupling expansion. We calculate high order results for the magnon
dispersions along the $b$-axis as well as the leading contribution along the 
$a$-axis. We discuss several mechanisms to explain the occurence of two low l
ying magnon branches without finding a profound supporting argument. 
We point out that symmetries of the magnon dispersions can be used to 
distinguish between classes of models with reflection and inversion symmetry. 
Further, we study the occurence of singlet states in the spin-cluster model in 
terms of the reported Raman observations. We find that there is no evidence for
a low lying singlet excitation of comparable energy to the lowest triplet 
excitation.

The Hamiltonian of the spin-cluster model reads
\begin{eqnarray}
H & = & J_1 \sum_n {\bf S}_{1,n}\cdot{\bf S}_{2,n}
      + J_2 \sum_n {\bf S}_{1,n}\cdot{\bf S}_{2,n+1} \nonumber \\
\label{H:cluster}
  & & + J'  \sum_n {(\bf S}_{1,n} + {\bf S}_{2,n}) \cdot 
		  ({\bf S}_{3,n} + {\bf S}_{4,n}) \\
  & & + J_3 \sum_n ({\bf S}_{2,n}\cdot{\bf S'}_{3,n} + 
	           {\bf S}_{4,n}\cdot{\bf S'}_{1,n} \nonumber \\
  & &\qquad\qquad + {\bf S}_{4,n}\cdot{\bf S'}_{2,n+1} +
	            {\bf S}_{1,n}\cdot{\bf S'}_{3,n+1}) \nonumber \;,
\end{eqnarray}
where the ${\bf S}_{i,n}$ and ${\bf S'}_{i,n}$ denote the four spins on the 
$n$'th cluster of two neighboring b-axis chains. 
$J_1=(1+\delta)J$ and $J_2=(1-\delta)J$ are alternating interactions along
the $b$-axis and $J_3$ is the interaction along the $a$-axis.
All interactions $J$, $J'$ and $J_3$ are assumed to be antiferromagnetic.

For an isolated cluster, $J_2=J_3=0$, we have two singlet, three triplet and 
one quintuplet eigenstates. We denote the low-lying eigenstates as follows:
\begin{eqnarray}
\psi_1 & = & {1\over\sqrt{3}}
  \left[ t_{12}^+t_{34}^- + t_{12}^-t_{34}^+ - t_{12}^0t_{34}^0\right] ~,
\nonumber \\
\psi_2 & = & s_{12}^{\phantom{0}} s_{34}^{\phantom{0}}~, \quad \quad
\psi_3^\alpha = s_{12}^{\phantom{0}}t_{34}^\alpha~,
\\
\psi_4^0 & = & {1\over\sqrt{2}}
\left[t_{12}^-t_{34}^+ - t_{12}^+t_{34}^-\right]~,
\nonumber
\end{eqnarray}
where $s_{ij}$ and $t_{ij}^\alpha$ are singlet and triplet states of the 
spins at sites $i$ and $j$ and $\psi_4^0$ is the $S^z = 0$ component of the
triplet $\psi_4^{\alpha}$. The corresponding energies are 
 $E_1 = -2J'+\frac{1}{4}J_1$, $E_2 = E_3 = -\frac{3}{4}J_1$ and
 $E_4 = -J'+\frac{1}{4}J_1$.

The ground state of the spin-cluster is the singlet state $\psi_1$ for 
antiferromagnetic couplings $J$ and $J'$ with $J' > \frac{1}{2} J_1$. 
For smaller values of $J'$ the ground state lies in the four-fold degenerate 
manifold of the states $\psi_2$ and $\psi_3^{\alpha}$.
In first order, the inter-cluster coupling $J_2$ lowers the energy of the 
 $\psi_3$ states while leaving the $\psi_2$ states unchanged. One thereby 
obtaines an effective $S=1$ Heisenberg chain with a Haldane gap at $\pi/(2b)$.

In the following we will focus on the first parameter regime where $\psi_1$
is the only groundstate. 
For an isolated cluster there are two low energy triplet excitations $\psi_3$ 
and $\psi_4$. To calculate their dispersion along {\bf b} we perform
a strong-coupling expansion \cite{Gelfand:90,Gelfand:96} around the isolated
cluster limit treating 
\begin{equation}
H_1 = J_2 \sum_n {{\bf S}_{1,n} \cdot {\bf S}_{2,n+1}}
\label{H1}
\end{equation}
as a perturbation. For the moment we neglect the inter-chain coupling $J_3$
which we assume to be significantly smaller than $J$ due to the longer
exchange path along {\bf a}.
We have calculated series up to order 7 in $J_2/J_1$ where the largest system
taken into account contains $L=32$ spins.

\begin{figure}
 \begin{center} 
 \epsfig{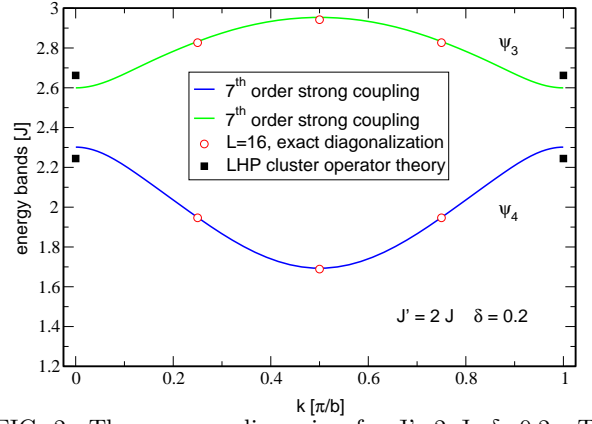}
 \caption[]
	{The magnon dispersion for J'=2~J, $\delta$=0.2. 
	 The solid lines are the result of the 7$^{\rm th}$ order strong 
	 coupling expansion, the open circles are exact diagonalization results
 	 and the squares denote the results from the cluster operator theory 
	 from \cite{Gros:00}.}
 \label{converged}
 \end{center}
\end{figure}

The two triplet dispersions are well seperated for large values of $J'$ as 
shown in Fig. \ref{converged}. Here the obtained series are very well 
converged already in second order. 
Our results are consistent with those of the reported exact diagonalization and
DMRG calculations, whereas the results of the linearized Holstein-Primakov 
approximation (LHP) used in the cluster-operator theory \cite{Gros:00}
disagree slightly for momenta in the region around $k=0$. 

\begin{figure}
 \begin{center} 
 \epsfig{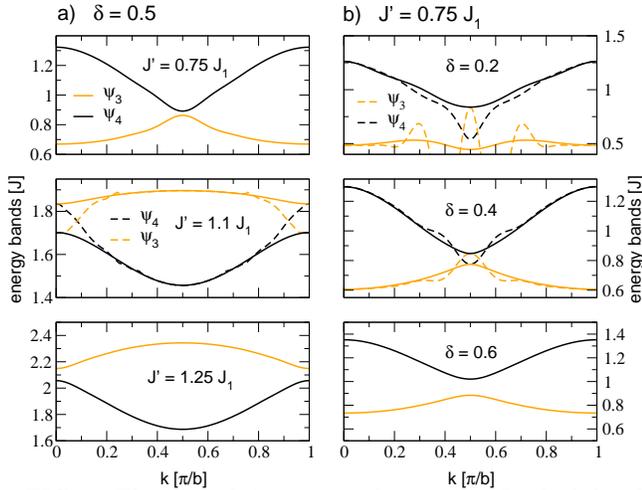}
 \caption[]
	{The $\psi_3$ and $\psi_4$ magnon dispersions. 
	 In the left column the cluster interaction $J'$ is varied for a 
	 constant dimerization $\delta$.
	 In the right column $\delta$ is varied keeping $J'$ constant.
	 A strong mixing of the two branches is observed.
	 Dlog Pad\'e approximants (dashed lines) have been used to extrapolate
 	 the series obtained treating the excitations seperately.
	 The solid lines are results of a combined calculation considering the
	 band repulsion. For $\delta = 0.2$ we find only a very limited 
	 convergence.}
 \label{Crossing}
 \end{center}
\end{figure}

The perturbative Hamiltonian $H_1$ strongly intermixes the two magnon 
excitations $\psi_3$ and $\psi_4$. For small values of the cluster exchange 
 $J'$ and the dimerization $\delta$ we find that the two energy bands are very 
close by for intermediate momenta $0 < k_y < \pi/(2b)$ as shown in Fig. 
\ref{Crossing}.

We have calculated effective Hamiltonians for the two excitations seperately
as well as a combined effective Hamiltonian treating the excitations on the 
same footing. We thereby consider the repulsive interaction between the two 
energy bands. 
While the first approach suffers a break-down of the perturbation expansion
due to small energy denominators in momentum regions where the two mixing 
magnon excitations are nearly degenerate, it allows to assign the nature
of the excitation at $k_y = \pi/(2b)$. 
For example, we find that for $J'=1.1~J_1$ and $\delta=0.5$ the lower energy 
at $k_y = 0$ corresponds to a $\psi_3$-excitation, whereas at $k_y = \pi/(2b)$ 
the lower excitation corresponds to a $\psi_4$-excitation. 
In Fig. \ref{Crossing} we have used Dlog Pad\'e approximants\cite{Baker} which 
extrapolate the calculated finite series to flatten out the occuring 
singularities (dashed lines).

The second approach allows to explicitely calculate the mixing between the
two excitation branches. We find that there is strong band repulsion for the
whole parameter range. For those parameter sets where in the first approach 
the extrapolated dispersions cross we now find two well separated branches, 
but the nature of the excitation changes depending on the momentum.

\begin{figure}
 \begin{center} 
 \epsfig{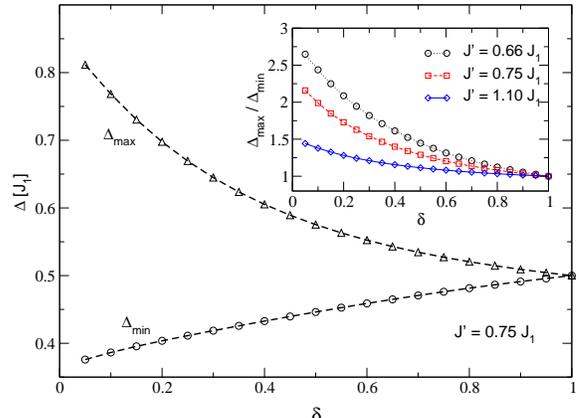}
 \caption[]
	{The maximum $\Delta_{max}$ of the $\psi_3$-excitation at 
         $k_y = \pi/(2b)$ and the gap $\Delta_{min}$ at $k_y = 0$ for 
	 $J'=0.75~J_1$ versus dimerization $\delta$.
	 The inset shows the ratio $\Delta_{max} / \Delta_{min}$ for varying 
	 values of $J'$ and $\delta$.
	}
 \label{Ratio}
 \end{center}
\end{figure}

Neutron scattering experiments observe the spin gap at the antiferromagnetic
point $k_y^{AF} = \pi/b$ and the zone center $k_y^{ZC} = 0$, which would
correspond to a $\psi_3$-excitation in the spin-cluster model. This assignment
enables us to determine the ratio of $\Delta_{max}$ at $k=\pi/(2b)$ to the gap
 $\Delta_{min}$ at $k=0$ in order to clarify wether an observed gap value
of around 10~meV is consistent with an estimate of 40 to 60~meV for the
spin exchange $J$.
In Fig. \ref{Ratio} we show the calculated values at the respective momenta
for $J' = 0.75~J_1$ as a function of dimerization $\delta$. For small values of
$\delta$ the results of the strong coupling expansion are less reliable, but
it seems that a ratio of 4 does not disagree with our results.
A tentatively estimated parameter set of $J'=0.66~J_1$ and $\delta = 0.05$
would give a ratio of around 3 and a spin gap of around $0.25~J_1$ which 
was doubted in a previous theoretical analysis \cite{Gros:00}.

Nevertheless, we point out that the shape of the $\psi_3$-dispersion is rather
flat in the vicinity of $k_y = 0$ and $k_y = \pi/b$ which is due to the strong 
band repulsion. This seems to contradict the experimental observations of a 
steep ascent \cite{Yosihama:98,Regnault:00}. 
Further we note that for these parameters the minimum of the 
$\psi_4$-excitation at $k_y = \pi/(2b)$ is of comparable size to the actual
spin gap, but has not been reported by neutron scattering experiments.

The leading contribution to the dispersion of the low-lying $\psi_3$-magnon
excitation along {\bf a} is given by the spin exchange mediated by an 
interaction $J_3$ between neighboring spin-clusters (see Fig. 
\ref{ClusterSpinModel}):
\begin{eqnarray}
  \epsilon_{ab}({\bf k_x}, {\bf k_y}) & = & \epsilon_b({\bf k_y}) \nonumber \\
   & & + \frac{J' \cdot J_3^2}{3J_1(J_1-2J')}
  	\cos({\bf k_x} \cdot {\bf a}) \cos({\bf k_y} \cdot {\bf b}) \;.
  \label{abDispersion}
\end{eqnarray}
The spin-cluster model inherently produces a leading periodicity of $2\pi/a$ 
for the dispersion along {\bf a}. This corresponds to the periodicity observed 
in INS experiments \cite{Yosihama:98,Regnault:00} and is consistent with a 
primitive unit cell of area $2ab$ as reported by structural x-ray 
investigations  \cite{deBoer:00} taking into account the $\pi/b$ periodicity 
along {\bf b}.
The $\psi_3$-magnon on an isolated cluster consists of a singlet state 
 $s_{12}$ along the $b$-axis bond. We therefore expect a strong coupling to 
scattering neutrons at the antiferromagnetic point $k_y^{AF} = \pi/b$ and only
a strongly suppressed signal in the vicinity of the zone center at 
 $k_y^{ZC} = 0$. According to (\ref{abDispersion}) the modulation of the magnon
dispersion along {\bf a} is out of phase for these choices of $k_y$.

Recent neutron scattering experiments \cite{Regnault:00} report the 
observation of two close by, but well separated, out of phase magnon branches.
The spin-cluster model at hand gives only one low lying magnon branch with the
observed periodicity. This excludes a simple spin-Peierls scenario to 
explain the occurence of $two$ out of phase magnon branches, e.~g. folding 
back a single magnon branch with double periodicity as it was suggested for a 
zig-zag model in \cite{Gros:99}.

A way of explaining the occurence of two close by magnon branches in this model
is to consider an anisotropic exchange interaction. 
An xxz-anisotropy could lift the triplet degeneracy in a single branch and a 
doublet branch. Experimentally this splitting of a single triplet branch 
should result in a 1:2 ratio of neutron scattering intensities which has not 
been reported \cite{Regnault:00}. 
Further, a magnetic field would cause the doublet line to split which has not 
been confirmed by optical spectroscopy \cite{Fischer:99}.

The investigation of the spin-cluster model reveals some new aspects
that should be covered by future neutron scattering experiments, namely the 
experimental evidence of asymmetric couplings and a verification of the basic
symmetries found in x-ray scattering.

Beside a careful investigation of the scattering intensities of the two magnon
branches, polarized neutron scattering experiments could give evidence for the 
splitting of a single magnon line, thereby proving the existence of anisotropic
couplings.

The verification of the basic symmetries by means of neutron scattering 
experiments is experimentally by far less sophisticated.
The spin-cluster model is symmetric under reflections along the mirror planes 
$\perp${\bf a} and $\perp${\bf b}. Accordingly, the obtained dispersion
 $\epsilon_{ab}({k_x}, {k_y})$ is symmetric under transformations
of ${k_x} \rightarrow -{k_x}$ and ${k_y} \rightarrow -{k_y}$.
The experimental data of the dispersion along {\bf a} for ${k_y} = 0$
and ${k_y} = \pi/b$ seem to be symmetric under the reflection 
${k_x} \rightarrow -{k_x}$. Nevertheless we point out that scanning
along an arbitrary value of ${k_y}$ will allow to distinguish between
models obeying double reflection symmetry and inversion symmetry. For the
latter we generally expect an unsymmetric dispersion for intermediate 
${k_y}$, whereas for models with double reflection symmetry we expect a
cosine modulation of the amplitude of a symmetric dispersion along {\bf a} as 
given in (\ref{abDispersion}).

For a single spin-cluster, there is one low-lying singlet excitation $\psi_2$ 
which is degenerate with the triplet excitation $\psi_3$. 
For vanishing inter-chain coupling $J_3$ the Hamiltonian (\ref{H:cluster}) is 
symmetric under a {\em local} interchange of ${\bf S}_{3,n}$ and 
 ${\bf S}_{4,n}$. The perturbation operator $H_1$ also conserves this local
symmetry. A direct product state of $\psi_1$ states which are even under this
symmetry and a $\psi_2$ excitation at some arbitrary cluster which is odd under
this symmetry couples to a variety of states with the same local symmetries, 
but does not allow the singlet excitation $\psi_2$ to move.
Though the energy of this excitation will be changed, it will not gain any 
dispersion. As a consequence the original degeneracy with the triplet 
excitation $\psi_3$ is lifted.
Neglecting the interaction along the $a$-axis we have calculated the energy of 
the singlet $\psi_2$ up to order 10 in $J_2/J_1$. 
It turns out that the ratio of the obtained energy to the gap of the 
elementary triplet excitation is very consistent with the expected value of 
1.6 for the $higher$ singlet excitation at 107~cm$^{-1}$ as measured in Raman 
spectroscopy. 

To explain the occurence of a low lying Raman excitation at 66~cm$^{-1}$  
(8.3~meV) \cite{Fischer:99} we have estimated the binding energies of 
2-magnon singlet bound states build of the elementary triplet excitations 
$\psi_3$ and $\psi_4$. We find that in leading order there is no 
substantial renormalization of these energies and we conclude that these bound
states are close to the 2-magnon continuum at $\sim$ 132~cm$^{-1}$ and are 
not sufficient to explain the occurence of the low lying Raman excitation.
Recent ESR studies show evidence for a Dzyaloshinskii-Moriya interaction below
20~K \cite{Luther:98,Nojiri:00}. The inclusion of spin-orbit coupling might
explain the observation of a Raman excitation degenerate to the spin gap.

In conclusion, we have calculated the magnetic excitations of a spin-cluster 
model. We find only partial agreement with the experimentally observed 
spectrum.
Along the crystallographic $b$-axis there are two strongly intermixing magnon 
bands in the parameter regime relevant for NaV$_2$O$_5$.  Due to strong band 
repulsion we find that the lower magnon branch exhibits a weaker dispersion
than experimentally observed. Further, the purely magnetic model leads to only
one magnon branch with the observed periodicity along the $a$-axis. 
We find one low lying singlet state which matches the energy of one of the 
observed Raman excitations.
While the prevailing models for the low temperature phase of NaV$_2$O$_5$ are 
inversion symmetric, the spin-cluster model obeys double reflection symmetry. 
To generally distinguish between these models we propose further experiments 
scanning the magnon dispersions along {\bf a} for arbitrary values of $k_y$.

We thank Girsh Blumberg for many valuable discussions. 
ST was supported in part by the German National Merit Foundation.

\end{document}